\documentclass[
    journal=jacsat,
    manuscript=article
]{achemso}

\usepackage[T1]{fontenc} % Use modern font encodings
\usepackage{siunitx}
\usepackage{physics}
\usepackage{ulem}
\usepackage[dvipsnames,table,xcdraw]{xcolor}
\definecolor{rmpblue}{HTML}{2e3092}
\usepackage[
	colorlinks=true,
	citecolor=rmpblue,
	linkcolor=rmpblue,
	filecolor=magenta,
	urlcolor=rmpblue,
]{hyperref}
\usepackage{booktabs} %for tables
\usepackage{tabularx}

\newcommand{\iu}{\mathrm{i}\mkern1mu} % imaginary unit
\newcommand{\eu}{\mathrm{e}\mkern1mu} % Euler number

\newcommand{\affilANU}{Research School of Physics, Australian National University, Canberra ACT 2601, Australia}

\newcommand{\affilSpain}{Departament of Physics, Universitat Politècnica de Catalunya - BarcelonaTech (UPC), Rambla Sant Nebridi 22, 08222 Terrassa (Barcelona), Spain}

\newcommand{\affilUK}{
 Institute of Photonics and Quantum Sciences, Heriot-Watt University, SUPA, Edinburgh EH144AS, United Kingdom}

\author{Eva Otero}
\altaffiliation{These authors contributed equally.}
\affiliation{\affilSpain}
\email{eva.otero@upc.edu}

\author{Ivan Toftul}
\altaffiliation{These authors contributed equally.}
\affiliation{\affilANU}
\email{ivan.toftul@anu.edu.au}

\author{Michael Scalora}
\affiliation{\affilUK}
\author{Jose Trull}
\affiliation{\affilSpain}
\author{Crina Cojocaru}
\affiliation{\affilSpain}
\author{Yuri Kivshar}
\affiliation{\affilANU}
\title[Probing topological edge states]{Probing topological edge states in multilayer dielectric photonic nanostructures}


\begin{document}

\begin{tocentry}

\includegraphics[width=\linewidth]{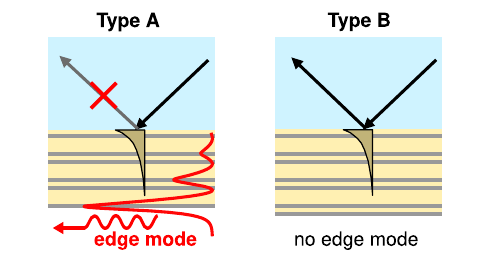}

\end{tocentry}

%%%%%%%%%%%%%%%%%%%%%%%%%%%%%%%%%%%%%%%%%%%%%%%%%%%%%%%%%%%%%%%%%%%%%
%% Abstract
%%%%%%%%%%%%%%%%%%%%%%%%%%%%%%%%%%%%%%%%%%%%%%%%%%%%%%%%%%%%%%%%%%%%%
\begin{abstract}
Topological photonics has recently attracted a significant attention, leading to many theoretical studies and predictions. However, direct experimental demonstrations of topology-driven effects, particularly in nanostructures, are rarely proven. In this work, we experimentally investigate multilayer photonic nanostructures composed of alternating subwavelength dielectric layers, arranged according to the topological Su-Schrieffer-Heeger model. By engineering different edge truncations, we probe different supported topological phases. Our theoretical predictions and experimental measurements reveal strong field confinement at the surface exclusively in the configuration supporting topological edge states, providing their localized character and topological origin. First, we demonstrate enhanced absorption in an ultrathin absorbing overlayer deposited on the surface when the topological surface mode is present, highlighting the potential for surface-sensing applications. Second, we observe enhanced second-harmonic generation in reflection from the surface, demonstrating the potential of topological states for surface enhanced nonlinear optics enabled by the strong localization of the electromagnetic field at the interface.
\end{abstract}

%%%%%%%%%%%%%%%%%%%%%%%%%%%%%%%%%%%%%%%%%%%%%%%%%%%%%%%%%%%%%%%%%%%%%
%% main part of the manuscript
%%%%%%%%%%%%%%%%%%%%%%%%%%%%%%%%%%%%%%%%%%%%%%%%%%%%%%%%%%%%%%%%%%%%%

\newpage

\section{Introduction}

One-dimensional (1D) photonic structures, consisting of a periodic variation of its optical properties, i.e. refractive index or optical thickness along a single direction, have been extensively used to control light propagation due to their geometric simplicity, relative ease of fabrication, and well-established theoretical framework. Light propagation can be effectively controlled through the engineering of the dispersion relation, enabling photonic bandgap formation, group-velocity reduction, and strong electromagnetic field localization. In particular, 1D photonic crystals can support strong electromagnetic field localization, making them attractive for increasing light–matter interactions, which leads to the enhancement of absorption or nonlinear effects when an active or nonlinear material is placed in the position of high intensity field. This property can be exploited in applications such as sensing, where the high field intensity could significantly increase device performance.

Notably, there is strong interest in the study of surface waves localized at the interface between photonic crystals and an external medium. In these modes, proposed years ago by Yariv and Yeh~\cite{Yariv1984}, the electromagnetic fields decay exponentially both inside the periodic medium and in the surrounding medium. In dielectric multilayers, they are referred to as Bloch surface waves (BSW) and have been extensively investigated for applications in optical sensing and surface-enhanced phenomena~\cite{Robertson1999, Liscidini2007, aurelio_electromagnetic_2017, doughan_temporal_2024}.
Although 1D periodic photonic crystals are a well established platform for controlling the light propagation, in the last years new ideas have appeared in the so-called topological photonics domain, a rapidly growing field that translates concepts originally developed in condensed matter physics and quantum mechanics into photonic systems~\cite{lu_topological_2014, Zhirihin2021Dec}. Topological photonics has attracted considerable attention for the potential of generating protected edge states at the interfaces between materials with distinct topological properties, inherently robust against perturbations. Although most studies have focused on two-dimensional systems~\cite{wang_observation_2009, hafezi_imaging_2013, kirsch_nonlinear_2021, Chen2021PRB}, there is also considerable research done in 1D geometries~\cite{Poshakinskiy2014PRL,Xiao2014PRX}.

Within this framework, the Su–Schrieffer–Heeger (SSH) model has become one of the best known examples of a 1D topological system. Originally introduced to describe electron transport in polyacetylene polymer~\cite{su_solitons_1979}, the SSH model describes a simple 1D lattice of two sites per unit cell, with alternating interaction energies (or coupling strength) J, and J’ between neighbouring sites. When J = J’, the system undergoes a topological phase transition. 
As a consequence, if we imagine a particle propagating through the lattice, it acquires a different Zak phase in each of the two topological regimes. This change in the topological invariant is responsible for the emergence of topologically protected edge states at the interface between regions with distinct topology~\cite{Atala2013NP}. Despite its conceptual simplicity, the SSH model has attracted considerable attention because it exhibits a rich variety of topological phenomena. More recently, it has also been successfully extended to photonic platforms, providing a versatile framework for investigating topological effects in light propagation and localization.

Photonic implementations of the SSH model have been successfully carried out in a variety of systems, including coupled waveguide arrays~\cite{Fu2020AS,malkova_observation_2009,zhang_experimental_2019}, ring resonators~\cite{zhao_topological_2018,Dobrykh2018PRL}, dielectric~\cite{Slobozhanyuk2015PRL, kruk_nonlinear_2019} and plasmonic~\cite{st-jean_lasing_2017,moritake_far-field_2022} zigzag chains, microwave photonic crystals~\cite{poli_selective_2015}, and polaritonic nanostructures~\cite{Orsini2024}. In contrast, comparatively few works have explored topological effects in multilayer structures. Existing studies proposing SSH-inspired geometries in multilayer structures are predominantly theoretical and have mainly focused on structures incorporating metals or two-dimensional materials, such as metal–vacuum stacks~\cite{jiang_topological_2021}, graphene multilayers~\cite{ge_topological_2015}, and metallic–dielectric stacks~\cite{deng_topological_2016}. To date, experimental demonstrations of topological surface states in multilayer systems remain scarce. Palatnik et al. realized a topological planar laser based on the interface between two dielectric 1D photonic crystals with different topological phases~\cite{Palatnik2021Nanophot}, but to the best of our knowledge, edge states in purely dielectric multilayer systems have not been reported.

In this work, we theoretically and experimentally investigate a 1D photonic crystal with an SSH-like geometry, establishing an all-dielectric platform for the realization and characterization of topological edge states. The alternating interaction energies of the SSH model are mimicked by varying the thickness of the dielectric layers consisting of nanometric size SiO$_2$ and Ta$_2$O$_5$. By selecting different edge truncations, we obtain nanostructures with different topological phases, one trivial and the other non-trivial, despite having the same periodicity. Only one of the geometries gives rise to topological edge states localized at the surface and appearing at frequencies within the bandgap [Fig.~\ref{fig:concept}]. We design the structure with the transfer-matrix method~\cite{wielian_transfer_2025}, excite the surface mode in the Kretschmann geometry, and confirm its surface character in two independent ways: a thin absorbing overlayer that drives the mode to critical coupling, and second-harmonic generation (SHG) in reflection. These findings demonstrate that multilayer photonic crystals provide a powerful all-dielectric platform in which topological phenomena can be associated with strong electromagnetic field confinement and enhanced linear and nonlinear light-matter interactions at optical interfaces.

\section{Topological surface states in multilayer structures}
\label{sec:theory}

\begin{figure}
    \centering
    \includegraphics[width=\linewidth]{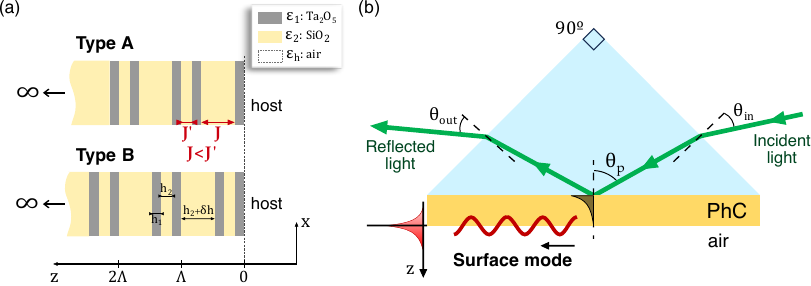}
    \caption{Concept. \textbf{(a)}~Two designs of the same SSH multilayer, consisting in two non-magnetic dielectrics ($(\varepsilon_1)$ and $(\varepsilon_2)$) alternating with period $\Lambda$. Type~A: cut at the weak bond (J). Type~B; cut at the strong bond (J'). \textbf{(b)}~Kretschmann configuration. The crystal is fabricated on a prism.}
    \label{fig:concept}
\end{figure}

\textbf{Problem formulation.} The structure is a stack of two non-magnetic dielectrics, Ta$_2$O$_5$ with $(\varepsilon_1, \mu_1{=}1)$ and SiO$_2$ with $(\varepsilon_2, \mu_2{=}1)$, with layer thicknesses $\{h_i\}$ and period $\Lambda = \sum_i h_i$ [Fig.~\ref{fig:concept}(a)]. By dimerizing the low-index spacer via alternating thicknesses $h_2$ and $h_2 + \delta h$ we map the multilayer onto the SSH chain: the high-index layers act as coupled waveguides, and the two spacer thicknesses set the weak and strong inter-waveguide bonds~\cite{Zhirihin2021Dec,Dobrykh2018PRL}. The mapping is exact in the waveguide regime $\beta > \sqrt{\varepsilon_2}\, k_0$, where the field is evanescent in the low-index spacers~\cite{wielian_transfer_2025}; the experiment operates close to this line, at $\beta \simeq 1.4\, k_0$ with $\sqrt{\varepsilon_2} = n_{\text{SiO}_2} = 1.45$. Here $k_0 = \omega/c$ is the vacuum wavenumber and $\beta$ the in-plane wavevector, conserved across all layers and the host. Truncating the semi-infinite stack at the weak bond leaves one high-index layer at the surface (Type~A), truncating at the strong bond leaves two (Type~B). The bulk is identical and only the termination differs.

The crystals are placed on top of a prism, H-K9L glass, and the light is coupled in Kretschmann configuration [Fig.~\ref{fig:concept}]. The total internal reflection at the prism base produces an evanescent wave with in-plane wavevector $\beta = n_{\text{p}} k_0 \sin\theta_{\text{p}}$, tunable through the angle $\theta_{\text{p}}$. On resonance it feeds the surface mode, which decays into both the crystal and the air.

\begin{figure*}
    \centering
    \includegraphics[width=\linewidth]{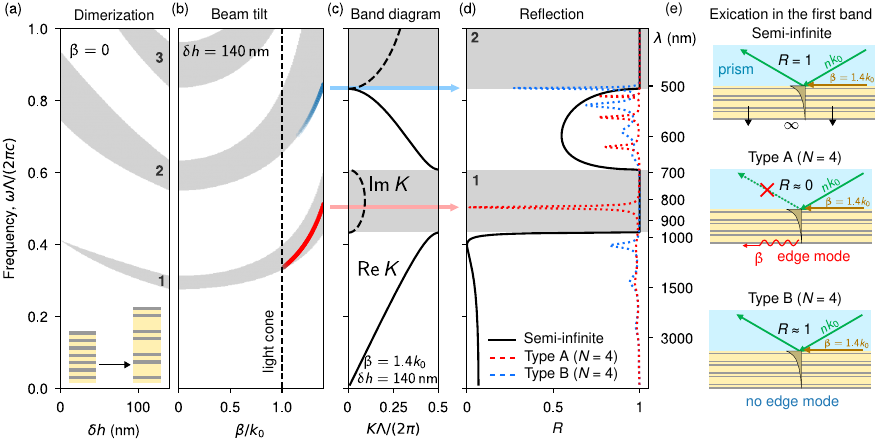}
    \caption{Design of the SSH surface state, from the T-matrix of Eqs.~\eqref{eq:dispersion}--\eqref{eq:r_inf}, TE polarization.
    \textbf{(a)}~Bulk bands (grey) versus dimerization $\delta h$ at normal incidence, $\beta = 0$. The three lowest gaps (labelled $1$--$3$) open as the low-index spacer dimerizes (inset).
    \textbf{(b)}~Bands versus the in-plane wavevector at $\delta h = 140$~nm. Right of the light cone $\beta = k_0$ (dashed) the host is evanescent and the terminations bind surface states, Eq.~\eqref{eq:disp_surf}: red in gap~1 (Type~A), blue in gap~2 (Type~B).
    \textbf{(c)}~Band diagram at the operating point $\beta = 1.4\,k_0$, $\delta h = 140$~nm. Solid: $\Re K$ in the bands; dashed: the decay $\Im K$ in the gaps (grey).
    \textbf{(d)}~Kretschmann reflectance $R = |r|^2$ for the same point. Black: semi-infinite crystal, Eq.~\eqref{eq:r_inf}, totally reflecting in the gaps. Red/blue dashed: finite Type~A and Type~B stacks of $N = 4$ periods. Only Type~A shows an edge-state dip in gap~1 ($\lambda \approx 800$~nm).
    \textbf{(e)}~Excitation cartoons for the three cases in (d).}
    \label{fig:bands}
\end{figure*}

\paragraph*{Band structure.}
We use the transfer matrix (T-matrix) in the full-wave basis $\varphi = (E_\perp, H_\perp)^{\mathsf T}$, continuous across every interface~\cite{Henriques2020PRA}. We do not fix the Cartesian components, since they differ for the transverse-electric (TE) and transverse-magnetic (TM) polarizations. Propagation across one period reads $\varphi(z+\Lambda) = \hat{T}_\Lambda\, \varphi(z)$ with
\begin{equation}
    \hat{T}_{\Lambda} = \prod\limits_{i=N}^{1}
    \begin{bmatrix}
        \cos\delta_i & \iu Z_i \sin\delta_i \\
        \iu Z_i^{-1}\sin\delta_i & \cos\delta_i
    \end{bmatrix},
    \label{eq:Tperiod}
\end{equation}
where $\delta_i = k_{iz} h_i$ is the phase across layer $i$, $k_{iz} = \sqrt{\varepsilon_i k_0^2 - \beta^2}$, and $Z_i$ is the transverse impedance of the layer~\cite[Sec.~7.2]{Orfanidis2002}: $Z_i^{\text{TE}} = Z_0\, \mu_i k_0 / k_{iz}$ or $Z_i^{\text{TM}} = Z_0\, k_{iz}/(\varepsilon_i k_0)$, with $Z_0 = \sqrt{\mu_0/\varepsilon_0}$. Bloch's theorem, valid for the semi-infinite crystal through the Ewald--Oseen extinction theorem~\cite[Sec.~2.4.2]{born2013principles,Yeh1977JOSA}, sets $\varphi(z+\Lambda) = \eu^{\iu K \Lambda}\varphi(z)$ and turns Eq.~\eqref{eq:Tperiod} into an eigenvalue problem for the Bloch vector $K$. With $\det\hat{T}_\Lambda = 1$ and no losses,
\begin{equation}
    K = \frac{1}{\Lambda}\cos^{-1}\!\left( \tfrac{1}{2}\tr\hat{T}_{\Lambda} \right).
    \label{eq:dispersion}
\end{equation}
Here we also used $\pm \cos^{-1} z = -\iu \ln \left( z \pm \iu \sqrt{1- z^2} \right)$ for any complex number $z$~\cite{abramowitz1964handbook}.
The trace is invariant under cyclic permutation, so $K(\omega)$ does not depend on where the crystal is cut. 
Hence Type~A and Type~B share the same bands, and they differ only in their surface response [Fig.~\ref{fig:bands} (a)-(d)].

\paragraph*{Surface modes.}
A surface state lives in a gap, defined by $\tfrac{1}{2}|\tr\hat{T}_\Lambda| > 1$, where $K = m\pi/\Lambda + \iu K''$ is complex ($m$ labels the gap). Choosing $K'' = \Im K > 0$ makes the Bloch part decay into the crystal as $\eu^{-K'' z}$ for $z > 0$. Binding the state to the surface also requires decay into the host, so the transverse field of the mode takes the form
\begin{equation}
    E(z) =
    \begin{cases}
        \alpha\, \eu^{\kappa z}\, \eu^{\iu\beta x}, & z \le 0 \quad (\text{host}), \\[2pt]
        E_K(z)\, \eu^{\iu K z}\, \eu^{\iu\beta x}, & z \ge 0 \quad (\text{crystal}),
    \end{cases}
    \label{eq:fullelectricfield}
\end{equation}
where $E_K(z+\Lambda) = E_K(z)$ is the periodic Bloch amplitude and $\kappa = \sqrt{\beta^2 - \varepsilon_{\text{h}} k_0^2}$ sets the decay into the host. The host field is evanescent, hence $\kappa$ real, only for $\beta > \sqrt{\varepsilon_{\text{h}}}\, k_0$, i.e. the mode must lie below the light line [Fig.~\ref{fig:bands}(b)]. Continuity of the transverse $E_{\perp}$ and $H_{\perp}$ at $z = 0$ fixes the amplitude $\alpha$ and gives the surface-mode dispersion,
\begin{equation}
    Z_{\text{h}} = \frac{T_{12}^{(\Lambda)}}{\eu^{\iu K\Lambda} - T_{11}^{(\Lambda)}},
    \label{eq:disp_surf}
\end{equation}
where $Z_{\text{h}} = Z_0\, \mu_{\text{h}} k_0/(-\iu\kappa)$ (TE) or $Z_0\,(-\iu\kappa)/(\varepsilon_{\text{h}} k_0)$ (TM) is the host transverse impedance. A root $(\omega,\beta)$ of Eq.~\eqref{eq:disp_surf} that falls inside a gap is a bound surface (Tamm/Bloch) wave~\cite{Martorell2006Jopt,Vinogradov2006PRB,Yeh1977JOSA}. The intra-crystal amplitude $E_K(z)$ is set by the eigenvector of $\hat{T}_\Lambda$ at the surface,
\begin{equation}
    \varphi_0 = \begin{pmatrix} T_{12}^{(\Lambda)} \\ \eu^{\iu K\Lambda} - T_{11}^{(\Lambda)} \end{pmatrix},
    \label{eq:phi0}
\end{equation}
taken with $\Im K > 0$: propagating $\varphi_0$ through the layer T-matrices of Eq.~\eqref{eq:Tperiod} reconstructs $E_K(z)$, hence the full field of Eq.~\eqref{eq:fullelectricfield}. The same T-matrix propagation through the finite prism-loaded stack yields the field profiles in Fig.~\ref{fig:cmt}(c,d).

\paragraph*{Which gap binds a state.}
Not every gap hosts a surface state. Whether a given termination binds one in a given gap is fixed by the topology of the bulk bands below it: the sum of their Zak phases, through the bulk--edge correspondence, counts the edge states in the gap~\cite{Zak1989PRL,Xiao2014PRX,wielian_transfer_2025}. 
For our parameters only Type~A binds a surface state in the first gap; the same gap of Type~B is empty. The correspondence is rigorous for continuous multilayers, and it protects the state by symmetry only when the termination cuts an inversion-symmetric unit cell~\cite{wielian_transfer_2025}. Our cuts fall at layer interfaces, so we verify the robustness of the surface mode against direct calculations.

To find the reflection coefficient of a semi-infinite photonic structure in terms of the T-matrix, one needs to consider the reflection scenario with the unit incident wave amplitude and reflected wave with amplitude $r$. 
On the boundary of the photonic crystal we have the state vector $\varphi(0) = (1+r , (1-r)/Z_{\text{h}})^T$, where $Z_{\text{h}}$ depends on the polarization (TE or TM). Next, Bloch theorem and T matrix formalism gives us $\varphi(\Lambda) = \hat{T}_{\Lambda} \varphi(0) = \eu^{\iu K \Lambda}\varphi(0)$, which finally leads to 
%The same states appear as the complex poles of the reflection coefficient of the semi-infinite crystal,
\begin{equation}
    r(\omega, \beta) = \frac{\eu^{\iu K\Lambda} - \left(T^{(\Lambda)}_{11} + Z_{\text{h}}^{-1} T^{(\Lambda)}_{12}\right)}{T^{(\Lambda)}_{11} - Z_{\text{h}}^{-1}T^{(\Lambda)}_{12} - \eu^{\iu K\Lambda}}.
    \label{eq:r_inf}
\end{equation}
The poles with $\Im\omega < 0$ are the \textit{radiative} edge states, i.e. leaky counterparts of the bound modes of Eq.~\eqref{eq:disp_surf}~\cite{Poshakinskiy2014PRL,wielian_transfer_2025}. We use reflection coefficient $R = |r|^2$ as the measured observable. Inside a gap the semi-infinite crystal reflects totally, $R = 1$ [Fig.~\ref{fig:bands}(d)].

The real samples are finite. A stack of $N$ periods sits between the prism substrate (impedance $Z_{\text{s}}$, the medium of incidence) and the air host ($Z_{\text{h}}$, at the free surface), with an optional extra surface layer (the Type~B termination or an absorbing top layer). Its reflectance follows from the total transfer matrix $\hat{T} = (\hat{T}_\Lambda)^N\, \hat{T}_{\text{surf}}$,
\begin{equation}
    r(\omega) = \frac{\frac{Z_\text{s}}{Z_\text{h}} T_{22}- T_{11} + Z_\text{s} T_{21} - Z_{\text{h}}^{-1}T_{12}}{\frac{Z_\text{s}}{Z_\text{h}} T_{22} + T_{11} - Z_\text{s} T_{21} - Z_{\text{h}}^{-1}T_{12}},
    \label{eq:r_finite}
\end{equation}
where $T_{ij}$ are the elements of $\hat{T}$ and $\hat{T}_{\text{surf}}$ is the T-matrix of the surface layer. Equation~\eqref{eq:r_finite} produces the Kretschmann spectra in Fig.~\ref{fig:bands}(d) and Fig.~\ref{fig:cmt}(c,d).

\paragraph*{Design.}
Figure~\ref{fig:bands} fixes the geometry. At normal incidence, dimerizing the spacer opens the low-order gaps [Fig.~\ref{fig:bands}(a)]. Tilting the beam raises the bands, and past the light cone $\beta = k_0$ each termination can bind a surface state: gap~1 for Type~A, gap~2 for Type~B [Fig.~\ref{fig:bands}(b)]. At the operating point $\beta = 1.4\,k_0$, $\delta h = 140$~nm the first gap sits at $\lambda \approx 800$~nm, with $\Re K$ in the bands and $\Im K$ across the gap [Fig.~\ref{fig:bands}(c)]. The reflectance confirms the selection: a finite Type~A stack shows an edge-state dip in gap~1 while Type~B does not, and the semi-infinite crystal reflects totally throughout [Fig.~\ref{fig:bands}(d,e)]. We therefore fix $h_1 = h_2 = 70$~nm, $\delta h = 140$~nm, and $\beta = 1.4\,k_0$: the two structures share the same period and gap, yet only the non-trivial Type~A carries a surface state near $\lambda = 800$~nm. A finite stack of $N = 4$ periods already localizes the field strongly enough and we could still resolve the spectral feature in the reflection, see also Fig.~\ref{fig:cmt}.

\section{Critical coupling in the Kretschmann geometry}
\label{sec:CMT}

To excite a surface state that lies below the air light line we use the Kretschmann geometry [Fig.~\ref{fig:concept}(b)]. The crystal sits on a prism; total internal reflection at the prism base launches an evanescent wave with in-plane wavevector $\beta = n_{\text{p}} k_0 \sin\theta_{\text{p}}$, propagating in the prism ($\beta < n_{\text{p}} k_0$) but evanescent in air. Sweeping the angle $\theta_{\text{p}}$ and the wavelength to the surface-mode dispersion feeds energy into the mode, which appears as a decrease (dip) in the reflected spectrum.

In this geometry the surface state is represented by a single resonance coupled to a single port. Temporal coupled-mode theory (TCMT) captures it~\cite{Hermann1984,Fan2003JOSAAJ,Suh2004IEEE}. The mode amplitude $a(t)$, normalized so that $|a|^2$ is the stored energy, obeys
\begin{equation}
\begin{split}
    \frac{\dd a}{\dd t} &= -\iu\omega_{\text{res}} a - (\gamma_{\text{rad}} + \gamma_{\text{abs}})\, a + \sqrt{2\gamma_{\text{rad}}}\, s^{+}(t), \\
    s^{-}(t) &= -s^{+}(t) + \sqrt{2\gamma_{\text{rad}}}\, a(t),
\end{split}
\label{eq:CMT}
\end{equation}
where $s^{\pm}$ are the incoming and outgoing wave amplitudes in the prism ($|s^\pm|^2$ are power fluxes), $\omega_{\text{res}}$ is the resonance frequency, $\gamma_{\text{rad}}$ is the leakage rate back into the prism, and $\gamma_{\text{abs}}$ collects all dissipation. Energy conservation, reciprocity, and time-reversal symmetry fix the coupling to $\sqrt{2\gamma_{\text{rad}}}$ and the direct background to $-1$~\cite{Fan2003JOSAAJ,Suh2004IEEE}.

\begin{figure*}
    \centering
    \includegraphics[width=\linewidth]{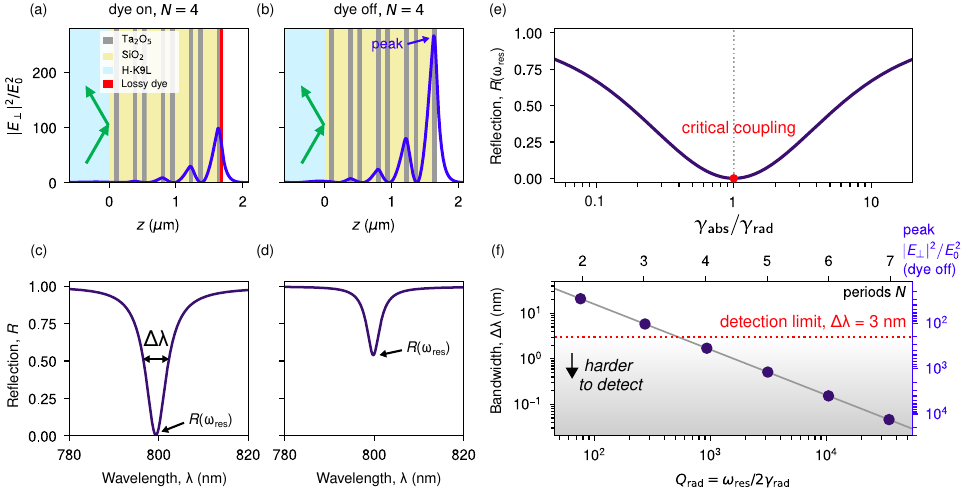}
    \caption{Critical coupling and the detection--enhancement trade-off.
    \textbf{(a,b)}~Field intensity $|E_\perp|^2/E_0^2$ through the $N = 4$ stack at $\lambda = 800$~nm, from prism to air, with (a) and without (b) the surface dye.
    \textbf{(c,d)}~Kretschmann reflectance of the stacks in (a) and (b). The dye deepens the dip to near zero (e), at critical coupling; the bare sample (d) keeps a shallow dip set by $\gamma_{\text{abs}}^{\text{fab}}$.
    \textbf{(e)}~On-resonance reflectance $R(\omega_{\text{res}})$ versus the loss ratio $\gamma_{\text{abs}}/\gamma_{\text{rad}}$, Eq.~\eqref{eq:Rmin}; the dip reaches zero only at critical coupling.
    \textbf{(f)}~Linewidth $\Delta\lambda$ (left) and peak field enhancement $|E_\perp|^2/E_0^2$ (right, dye off) versus the radiative quality factor $Q_{\text{rad}} = \omega_{\text{res}}/2\gamma_{\text{rad}}$ and the number of periods $N$. Dots: T-matrix poles; grey line: the CMT tunneling law.
    Parameters: TE polarization, resonance at $\lambda = 800$~nm; unit cell SiO$_2$($70$)\,/\,Ta$_2$O$_5$($70$)\,/\,SiO$_2$($210$)\,/\,Ta$_2$O$_5$($70$)~nm; $\varepsilon_{\text{prism}} = 2.28$, $\varepsilon_{\text{SiO}_2} = 2.11 + \iu\,0.001$, $\varepsilon_{\text{Ta}_2\text{O}_5} = 4.46$, $\varepsilon_{\text{dye}} = 2.18 + \iu\,0.5$.}
        \label{fig:cmt}
\end{figure*}

The two rates are set differently. The radiative rate is fixed by tunneling through the $N$ periods between the prism and the surface, $\gamma_{\text{rad}} \propto \eu^{-2\Im(K)\Lambda N}$, so it is a design knob [Fig.~\ref{fig:cmt}(b)]. The dissipative rate splits as $\gamma_{\text{abs}} = \gamma_{\text{abs}}^{\text{fab}} + \gamma_{\text{abs}}^{\text{dye}}$: a floor $\gamma_{\text{abs}}^{\text{fab}}$ from fabrication (roughness, thickness errors, residual absorption), plus a tunable term $\gamma_{\text{abs}}^{\text{dye}}$ from a thin surface dye.

For a monochromatic pump, $s^{+}(t) = s^{+} \eu^{-\iu \omega t}$, the steady state of Eq.~\eqref{eq:CMT} gives the energy build-up and the reflectance, [$R(\omega) = |r(\omega)|^2 = |s^{-} / {s^{+}}|^2$],
\begin{equation}
    \frac{|a|^2}{|s^{+}|^2} = \frac{2\gamma_{\text{rad}}}{(\gamma_{\text{rad}} + \gamma_{\text{abs}})^2 + (\omega - \omega_{\text{res}})^2},
    \label{eq:buildup}
\end{equation}
\begin{equation}
    R(\omega) = 1 - \frac{4 \gamma_{\text{abs}} \gamma_\text{rad}}{\left( \gamma_{\text{abs}} + \gamma_{\text{rad}} \right)^2 + \left( \omega - \omega_{\text{res}} \right)^2}, \qquad R(\omega_{\text{res}}) = \left( \frac{\gamma_{\text{rad}} - \gamma_{\text{abs}}}{\gamma_{\text{rad}} + \gamma_{\text{abs}}} \right)^{2}.
    \label{eq:Rmin}
\end{equation}
Far off resonance $R \to 1$ (total internal reflection). Without absorption, $\gamma_{\text{abs}} = 0$, Eq.~\eqref{eq:Rmin} gives $R = 1$ at \textit{all} frequencies: the mode is excited, yet all stored energy tunnels back coherently and only the phase of $r$ winds by $2\pi$. A dip in reflection therefore requires loss. It reaches zero at \textit{critical coupling} $\gamma_{\text{abs}} = \gamma_{\text{rad}}$, where leakage and reflection interfere destructively [Fig.~\ref{fig:cmt}(e)]. The depth measures the ratio $\gamma_{\text{abs}}/\gamma_{\text{rad}}$ but is symmetric under $\gamma_{\text{abs}} \leftrightarrow \gamma_{\text{rad}}$; a small known added loss lifts the ambiguity (the dip deepens if $\gamma_{\text{abs}} < \gamma_{\text{rad}}$).

This sets the two roles of the loss. Adding absorption at the surface (e.g. a thin absorbing overlayer) raises $\gamma_{\text{abs}}$ toward $\gamma_{\text{rad}}$, deepening the dip; because the absorption is only at the surface, a dip that deepens on this overlayer certifies that the field sits at the surface, not inside the stack [Fig.~\ref{fig:cmt}(a)--(d)]. The bare sample already shows a shallower dip, which through Eq.~\eqref{eq:Rmin} measures the fabrication floor $\gamma_{\text{abs}}^{\text{fab}}$. The dip is a detection tool, not a field gauge: its depth tracks $\gamma_{\text{abs}}/\gamma_{\text{rad}}$, while the stored energy \textit{decreases} as $\gamma_{\text{abs}}$ grows [see Eq.~\eqref{eq:buildup} and Fig.~\ref{fig:cmt}(c,d)]. The same absorption that deepens the dip weakens the field. We therefore detect the mode with the absorbing layer, and for the nonlinear measurement remove it, leaving the hotspot $|a|^2/|s^+|^2 = 1/(2\gamma_{\text{abs}}^{\text{fab}})$ capped only by the fabrication floor.

Figure~\ref{fig:cmt}(f) shows that when the number of periods (N) increases both the quality factor $Q_{\text{rad}} = \omega_{\text{res}}/2\gamma_{\text{rad}}$ and the peak field enhancement (at the surface of the structure) $|E_\perp|^2/E_0^2$ raises. From these results one would expect that when fabricating a real finite stack sample the more number of periods the better. However, with each period added the line width of the resonance $\Delta\lambda$ narrows and below $\Delta\lambda \approx 3$~nm the mode gets harder to couple experimentally. For this reason, in the following experimental study we chose a multilayer structure with $N = 4$ periods, representing a compromise between strong field localization, a sufficiently high quality factor, and the practical constraints associated with experimental detection.

To assess the robustness of the proposed design, random perturbations were introduced into the thicknesses of all layers of our design, and the resulting angular shift of the surface mode in the Type~A structure at 800~nm was evaluated. Two levels of disorder were considered ($\pm 2\%$ and $\pm 10\%$), with each layer assigned a randomly varied thickness within the corresponding range around its nominal value, thereby generating non-ideal, non-periodic structures (see Supporting Information S1 for details). For comparison, the same analysis was performed on a conventional periodic bilayer structure composed of the same materials and also supporting a surface state.

The results demonstrate that the proposed topological structure is more robust against moderate thickness variations. Notably, for the conventional periodic bilayer structure with $\pm 10\%$ disorder, several randomly generated configurations no longer supported a surface state. In contrast, the Type~A structure consistently preserved the surface mode under the same level of perturbation. These findings highlight the greater tolerance of the proposed design to fabrication imperfections and confirm its superior robustness compared to conventional periodic multilayer structures.

\section{Topology-driven localization of light}
\label{sec:linear}

Once the topology-driven light localization at the surface has been theoretically predicted and analysed, we aim to obtain a direct experimental demonstration of this effect by measuring the light localization in the superficial layer. Such measurements are rare and challenging, as the SSH modes are not directly accessible since they strongly decay within the propagation in air. To investigate them experimentally, two 1D structures (Type~A and Type~B) were fabricated directly on top of a high index glass prism, as described in Fig.~\ref{fig:cmt}(a,b). Both structures share the same unit cell, consisting of SiO$_2$($70$)\,/\,Ta$_2$O$_5$($70$)\,/\,SiO$_2$($210$)\,/\,Ta$_2$O$_5$($70$)~nm; $\varepsilon_{\text{prism}} = 2.28$, $\varepsilon_{\text{SiO}_2} = 2.11 + \iu\,0.001$, $\varepsilon_{\text{Ta}_2\text{O}_5} = 4.46$, $\varepsilon_{\text{dye}} = 2.18 + \iu\,0.5$.

A TE-polarized supercontinuum laser beam was focused onto the side of the high index prism carrying the multilayer, following the sketch in Fig.~\ref{fig:concept} (b). The pulse duration is long enough to ensure adequate spectral resolution of the resonances induced by the surface states. The incidence angle $\theta_{\text{in}}$, was controlled using a rotation stage. From this angle, the internal prism angle $\theta_p$ and corresponding in-plane wavevector, $\beta = n_{\text{p}} k_0 \sin\theta_{\text{p}}$, were determined. The reflected light was subsequently collected and measured with a spectrometer. Further details regarding the experimental setup are provided in the Supporting Information S2.

\begin{figure}[H]
    \centering
    \includegraphics[width=\linewidth]{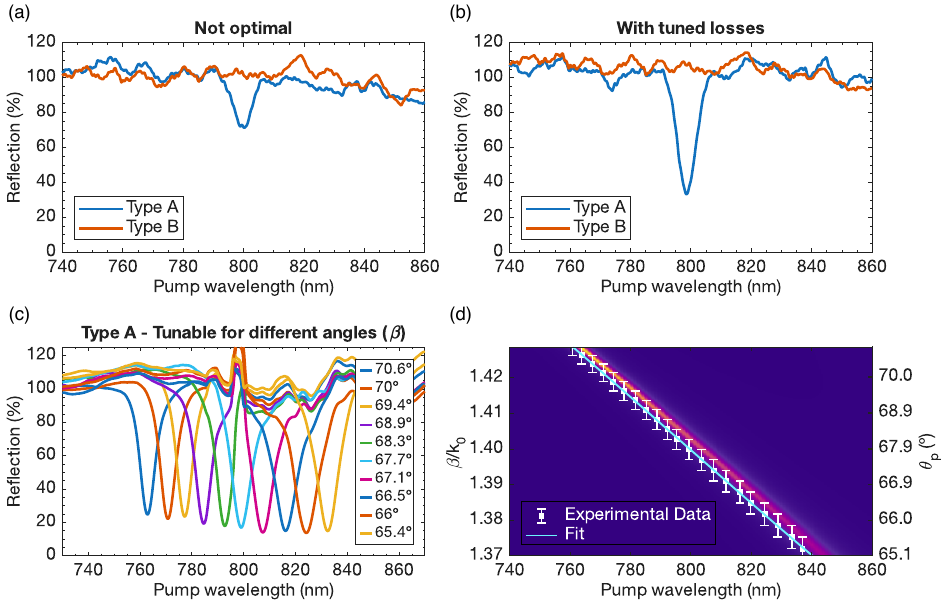}
    \caption{Kretschmann reflectance measurements, TE polarization.
    \textbf{(a,b)}~Reflection at $\theta_{\text{p}} \simeq \ang{67.5}$ of the bare samples (a) the samples with the dye coating (b).
    \textbf{(c)}~Reflection spectra close to the surface mode for different incident angles: the resonance tunes continuously with $\theta_{\text{p}}$ (i.e. with $\beta$).
    \textbf{(d)}~Computed vs experimental measure of the surface-mode dispersion as a function of wavelength, wavevector $\beta /k_0$ and angle $\theta_{\text{p}}$; markers track the resonances measured in (c) with the corresponding error bars.}
    \label{fig:exp}
\end{figure}

Figure~\ref{fig:exp} shows the spectral dependence of the measured Kretschmann prism reflectance, $R = |r|^2$, as a function of the incident angle. Type~A sample exhibits a shallow dip near $\lambda = 800$~nm, whereas the reflectance of Type~B remains essentially flat [Fig.~\ref{fig:exp}(a)]. As theoretically predicted, only the non-trivial termination (Type~A) exhibits a bound surface state. The depth of the reflectance dip provides a measure of the fabrication-induced loss floor, $\gamma_{\text{abs}}^{\text{fab}}$, defined in Eq.~\eqref{eq:Rmin}.
To selectively increase losses at the interface, the surface od the photonic crystal was coated with an absorbing dye (Q-Switch I) whose absorption band overlaps the resonance wavelength near 800~nm (details in the Supporting Information S3). The coating substantially deepens the Type~A reflectance dip, bringing the resonance closer to critical coupling, while producing no noticeable change in the Type~B structure [Fig.~\ref{fig:exp}(b)]. Since the dye is confined to the surface, this selective enhancement provides direct experimental evidence that the resonance is a surface-localized state rather than a mode distributed within the multilayer, consistent with the model described by Eq.~\eqref{eq:Rmin} and Fig.~\ref{fig:cmt}.

The reflection spectra measured at a series of angles (corresponding to the samples coated with the absorbing dye) shows the angular and spectral dependence of the surface mode [Fig.~\ref{fig:exp}(c)].
The numerically computed surface-mode dispersion map in Fig.~\ref{fig:exp}(d) shows the continuous tuning of the resonance as a function of the parameter $\beta$. The experimentally measured resonance positions closely follow the calculated dispersion, showing excellent agreement between experiment and simulation. The observed shift between the two branches is on the order of $\beta \sim 0.002-0.003\,k_0$, equivalent to $\theta_p \sim 0.24-\ang{0.3}$. Such a deviation is fully consistent with the fabrication tolerances and refractive-index uncertainties of the fabricated multilayer structures.

Type~B and the TM channel show only total internal reflection (Supporting Information S4), consistent with the band structure of Fig.~\ref{fig:bands}.

\section{Topology-enable SHG generation }
\label{sec:SHG}

Topology-driven surface light localization can also be probed via second-order nonlinear optical processes arising from the strong electromagnetic field confinement at the surface of the outermost layer. As both Ta$_2$O$_5$ and SiO$_2$ are centrosymmetric materials, the bulk electric-dipole contribution to the second-order susceptibility is forbidden by symmetry. Consequently, SHG can only originate from the interfaces (Coulomb contribution) and from the magnetic Lorentz force term, $\vb{J}\times\vb{B}$, of the nonlinear wave equation. The resulting second-order nonlinear polarization, $\vb{P}^{(2)}$, is therefore dominated by these surface and magnetic contributions. Under a TE pump the field has no longitudinal component ($E_z = 0$), so the surface Coulomb term is absent and the magnetic Lorentz term is the sole mechanism; it converts the TE pump into a TM second harmonic \cite{scalora_second-_2010}. Although this contribution is intrinsically weak, it is governed by the local electric and magnetic fields, which are simultaneously enhanced and co-localized by the surface mode at the outermost Ta$_2$O$_5$ layer. The predicted field intensity enhancement at the surface mode is of the order of 300, effectively enhancing the TE-driven SHG process.

\begin{figure}
    \centering
    \includegraphics[width=\linewidth]{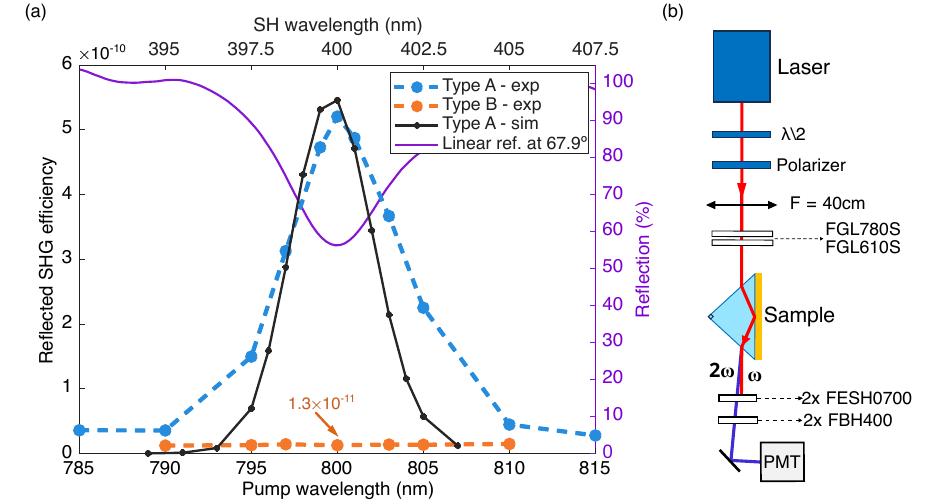}
    \caption{Second-harmonic generation in reflection. \textbf{(a)}~Reflected SHG efficiency versus pump wavelength at $\theta_{\text{p}} = \ang{67.9}$. Experimental SHG measurements for Type~A (blue) and Type~B (orange); Simulation results (black line); Type~A reflection with dye at the surface (purple, right axis). \textbf{(b)}~Setup configuration: the harmonic is selected by spectral filters and detected by a photomultiplier.}
    \label{fig:shg}
\end{figure}

The SHG signal was measured in reflection from the bare Type~A and Type~B samples (without the dye). A fixed incidence angle of $\theta_{\text{p}} = \ang{67.9}$ was chosen, corresponding to a surface-mode resonance centered near 800~nm [Fig.~\ref{fig:shg}(a); purple curve, right axis]. The samples were pumped by a tunable femtosecond laser source (180 fs pulse duration, 9~GW/cm\textsuperscript{2} peak intensity). The generated SHG signal was isolated using spectral filters and detected with a PMT [Fig.~\ref{fig:shg}(b)].
The reflected SHG efficiencies measured for the structure Type~A (blue line) and Type~B (red line) are shown in Figure~\ref{fig:shg}(a). 
The efficiency of Type~A sample peaks sharply at the surface mode resonance ($\lambda_{\text{pump}} = 800$~nm), with a peak efficiency of $\sim\!5\times10^{-10}$, while Type~B sample efficiency stays weak ( $\sim\!1.3\times10^{-11}$) and featureless in wavelength. The background contribution from the prism by itself has been checked and found to be negligible. The experimental results are compared with simulations of the SHG predictions. 
We implemented a hydrodynamic-Maxwell model that incorporates the full linear and nonlinear dispersive response of the two layers materials and explicitly accounts for bulk, surface, and magnetic nonlinear contributions, without imposing an effective second-order susceptibility. The calculated SHG efficiency, generated predominantly at the surface of the outermost layer, closely follows the spectral dependence of the surface-mode resonance, reaching a maximum at 800~nm. The predictions, shown as continuous black line in Fig.~\ref{fig:exp}(a) are in a very good agreement with the measurements. More details about the theoretical model can be found in references~\citenum{scalora_dynamical_2012,scalora_nonlinear_2015}.

These results confirm that the SHG enhancement is resonant and occurs only in the Type~A structure, where the non-trivial truncation supports strong surface-field localization. The close correspondence between the SHG spectrum and the surface-mode resonance demonstrates that the enhanced second-harmonic generation is directly driven by resonant optical-field confinement at the surface.

\section{Conclusions}

We have theoretically and experimentally demonstrated strong light localization in multilayer photonic nanostructures composed of alternating subwavelength dielectric layers arranged according to the topology of the SSH model. By investigating two distinct edge terminations, we have shown that the truncation alone determines whether a topological surface state emerges. Both numerical simulations and experimental measurements reveal that only the non-trivial termination supports a surface-bound mode, characterized by strong co-localization of the electric and magnetic fields at the interface. Using the Kretschmann configuration, we experimentally mapped the angular and spectral dispersion of the surface mode and found excellent agreement with theoretical predictions. By coating the surface with an ultrathin absorbing dye layer, we drove the resonance to critical coupling and observed enhanced absorption exclusively in the presence of the topological surface mode, demonstrating its potential for highly sensitive surface-sensing applications. As a second experimental verification, we measured resonantly enhanced second-harmonic generation in reflection, providing additional evidence of the strong field localization associated with the topological state. 

These results establish dielectric multilayer nanostructures as a platform for the investigation of topological surface states with enhanced light-matter interactions. The strong electromagnetic confinement achieved at the interface opens promising opportunities for nanoscale surface-sensitive linear and nonlinear optical applications, including sensing, spectroscopy, biotechnology, and environmental monitoring.

%%%%%%%%%%%%%%%%%%%%%%%%%%%%%%%%%%%%%%%%%%%%%%%%%%%%%%%%%%%%%%%%%%%%%
\begin{acknowledgement}
EO acknowledges the support of the predoctoral program AGAUR-FISDUR (2023 FISDU 00050) of the Secretariat of Universities and Research of the Department of Research and Universities of the Generalitat of Catalunya and the European Social Plus Fund.
IT and YK were supported by the Australian Research Council (Discovery project DP210101292) and the International Technology Center Indo-Pacific (ITC-IPAC) via Army Research Office (contract FA520923C0023). 
JT, and CC acknowledge the Spanish Agencia Estatal de Investigación from Ministerio de Ciencia, Investigación y Universidades (project PID2023-148620NB-I00 granted by MICIU/AEI/10.13039/501100011033 and FEDER, UE). 

\end{acknowledgement}

%%%%%%%%%%%%%%%%%%%%%%%%%%%%%%%%%%%%%%%%%%%%%%%%%%%%%%%%%%%%%%%%%%%%%
\begin{suppinfo}

The Supporting Information contains further details on the robustness study, experimental setup, absorbing dye, linear experimental measurements and extra theory details.

\end{suppinfo}

\newpage 
%%%%%%%%%%%%%%%%%%%%%%%%%%%%%%%%%%%%%%%%%%%%%%%%%%%%%%%%%%%%%%%%%%%%%
\bibliography{references}

\end{document}